\documentclass[pra,twocolumn,showpacs]{revtex4}
\usepackage{graphicx}
\begin{document}

\title{The Impossibility of the Counterfactual Computation for all
Possible Outcomes.}

\author{L. Vaidman}
\affiliation{ School of Physics and Astronomy\\
Raymond and Beverly Sackler Faculty of Exact Sciences \\
Tel-Aviv University, Tel-Aviv 69978, Israel}

\date{\today}

\begin{abstract}
Recent proposal for counterfactual computation [Hosten et al.,
Nature, 439, 949 (2006)] is analyzed. It is argued that the method
does not provide counterfactual computation for all possible
outcomes. The explanation involves a novel paradoxical feature of
pre- and post-selected quantum particles: the particle can reach a
certain location without being on the path that leads to this
location.
\end{abstract}


\maketitle


The foundations of quantum theory which is
 almost 100 years old are
still a subject of a heated debate. Where is a quantum particle
passing through a Mach-Zehnder interferometer? Is it present in two
places simultaneously? If not, how to explain the interference at
the output? Recently, Hosten et al. \cite{HON} claimed that quantum
mechanics allows to build another counterintuitive device: a
computer which yields the outcome of a computation without running.
It is an improvement of the proposal of Jozsa \cite{Jo}, who used an
interaction-free measurement (IFM) scheme \cite{IFM} for
constructing a computer which could give the outcome without
running, but only in the case of one particular outcome.

It was claimed \cite{MIFM} that the original IFM can find that there
{\em is} an object in a particular place in an interaction-free
manner, but it cannot find that the place is {\em empty} in the
interaction-free way. Also, Jozsa and Mitchison \cite{Jo,MJ1,MJ2}
argued that a computer cannot yield all possible outcomes without
running. Hosten et al. provided an apparent counterexample to these
results. I will argue, that one cannot claim that their device
provides all possible outcomes without running, but the way it fails
to do so yields a novel insight into quantum behavior.

 The  Hosten et al. proposal is based on
``chained'' Zeno effect improvement \cite{Kw} of the IFM method. A
simple way to describe it is as follows. We have a computer which is
a device that performs computation when a photon enters from an
input port. There are only two possible outcomes of the computation,
0 and 1. If the outcome is 0, the photon passes through without any
disturbance (or with a known delay which can be compensated). If the
outcome is 1, the photon is absorbed by the device.

First, let us describe the original counterfactual computation (CFC)
proposal \cite{Jo}. Two identical optical cavities  are connected by
a common wall - an almost 100\% reflection mirror. During the period
of the oscillation in a cavity, a photon starting in the left ($L$)
cavity evolves into a superposition of being in the left and in the
right ($R$) cavities according to the following law:
\begin{eqnarray}
\label{U} \nonumber
|L\rangle &\rightarrow & \cos \alpha  |L\rangle + \sin \alpha  |R\rangle,\\
 |R\rangle &\rightarrow & -\sin \alpha  |L\rangle + \cos\alpha
 |R\rangle, ~~~~~~~~~~~\alpha \ll 1.
\end{eqnarray}
 The
computer device is placed inside the right cavity, see Fig. 1. A
single photon localized wavepacket starts in cavity $L$ and it is
left to evolve $N$ periods of the oscillation of a single cavity. It
is arranged that $N\alpha={\pi \over 2}$.
\begin{figure}[b]
  \includegraphics[width=5cm]{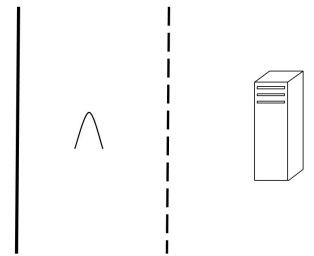}\\
\caption{The Jozsa counterfactual computation method. If the outcome
is 1, the photon remains in the left cavity, and if the outcome is
0, the photon moves to the right cavity.  } \label{1}
\end{figure}

 Assume first that the outcome is 0. After one period, the state of
the photon is $~\cos \alpha  |L\rangle + \sin \alpha |R\rangle$. It
is easy to see that after $n$ periods, the state is $~\cos n\alpha
|L\rangle + \sin n\alpha  |R\rangle$.  When the number of periods
equal $N={\pi \over{2\alpha}}$, the final state of the photon is
 $|R\rangle$.

Now assume that the outcome is 1. After one period, the state of the
photon is $\cos \alpha  |L\rangle + \sin \alpha |abs~ 1 \rangle$.
The state $|abs~ 1 \rangle$ signifies the photon absorbed by the
computer device on the first round. (The process might include
macroscopic amplification,  in which case the situation is usually
described as a collapse to the state $|L\rangle $ with probability
$\cos^2 \alpha$, or collapse to the absorption of the photon with
probability $\sin^2 \alpha$ .) After $N$ periods the state is
\begin{equation}\label{Nperi}
    \cos^N \alpha  |L\rangle + \sum_{n=1}^N \cos^{n-1}\alpha ~\sin \alpha
|abs~ n \rangle .
\end{equation}

The final step of the procedure is a measurement which tests the
presence of the photon in cavity $L$. If we find the photon there,
we know that the outcome is 1, since, if the outcome is 0, after $N$
periods it has to be in cavity $R$. And we know that it was computed
counterfactually, since computation of the outcome 1 ends by the
absorption of the photon by the computing device, but the photon was
detected by another detector.

If we do not find the photon in cavity $L$, we know that there is a
high probability for the outcome is 0, but there is no any reason to
claim that in this case there have been a counterfactual
computation, since at every period part of the photon wave went
through the computer device.

The probability to absorb the photon by the computer in case the
outcome is 1 is  $ 1-\cos^{2N}{\pi \over 2N}\approx {\pi ^2 \over
4N}$, so we have constructed a CFC procedure for a single outcome
with efficiency which can be  arbitrary  close to 100\%.

In order to perform a counterfactual computation for both outcomes,
Hosten et al. proposed to construct three identical optical cavities
using two identical almost 100\% reflection mirrors. The computer
device is placed inside the third cavity, see Fig. 2.
\begin{figure}[b]
  \includegraphics[width=6cm]{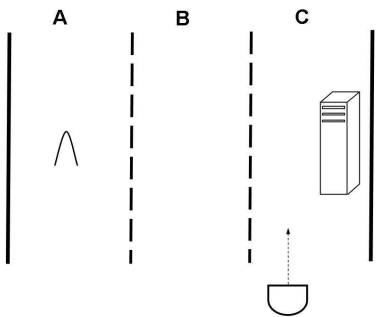}\\
  \caption{Hosten et al.  counterfactual computation method. If the outcome is
0, the photon remains in  cavity $A$, and if the outcome is 1, the
photon moves to cavity $B$. } \label{2}
\end{figure}

 Again, a single photon localized wavepacket
starts in cavity $A$ and moves toward the  almost 100\% reflection
mirror which separates cavity $A$ from cavity $B$. After it bounces
on the mirror transmitting  a localized packet of an amplitude $\sin
\alpha$ to the cavity $B$,  the mirror between cavities $A$ and $B$
is switched to be 100\% reflective. It remains to be 100\%
reflective for $N$ periods. During this time, the part of the photon
wave in cavity $A$ remains as it is, and the part in cavity $B$
evolves as in the previous two-cavity case. Thus, if the outcome is
0, the state of the photon becomes
\begin{equation}\label{Nperi0}
    \cos \alpha  |A\rangle + \sin \alpha |C\rangle ,
\end{equation}
and if the outcome is 1, the state of the photon becomes
\begin{equation}\label{Nperi1}
\cos \alpha  |A\rangle +  \sin \alpha \left (\cos^N \alpha |B\rangle
+ \sum_{n=1}^N \cos^{n-1}\alpha ~\sin \alpha |abs~n \rangle \right
).
\end{equation}

At this stage a measurement is performed which tests the presence of
the photon in cavity $C$. The probability to find the photon in this
measurement does not vanish only if the outcome is 0, but even then
it is very small: $p_{fail}\simeq {\pi^2\over 4N^2}$. If we find the
photon, we know the outcome of the computation, but CFC fails, since
the photon did pass through the computer.
 If the outcome is 1, we
have even  smaller  probability of the CFC failure, i.e. absorption
of  the photon by the computer device.
 If the photon was not
absorbed, then the state of the photon in case of the outcome 0 is
just $|A\rangle$, and in case of the outcome 1, the state is, up to
normalization, $\cos \alpha |A\rangle + \sin \alpha
 \cos^N \alpha |B\rangle$. This ends the first subroutine of the
 computation process.

 Now, the mirror between cavities
$A$ and $B$ is opened again for one bounce of the photon, i.e, it
leads to the evolution described by (\ref{U}), after which it is
closed for $N$ periods as before. The second run of the subroutine,
 as the first one, ends by the test of the presence of the photon
in $C$. If the outcome is 0, everything is the same as in the first
subroutine: probability of the failure of the CFC is  again
$~p_{fail}\approx {\pi^2\over 4N^2}$ and if the photon is not found
in $C$, its final state is $|A\rangle$. If the outcome is 1, we have
a tiny probability for a failure and if the photon is not absorbed
by the computer, the final state of the photon is, up to
normalization, $~(\cos^2 \alpha- \sin^2 \alpha \cos^N  \alpha)
|A\rangle + (\sin \alpha \cos^{N+1} \alpha + \sin \alpha \cos^{2N+1}
\alpha) |B\rangle$.
  The state is approximately equal
 to  $~\cos 2\alpha  |A\rangle +
 \sin 2\alpha |B\rangle$.  After $N$ rounds the state is approximately equal
 to  $\cos N\alpha  |A\rangle +
 \sin N\alpha |B\rangle ~= |B\rangle$.

 Under this approximation, after $N$ rounds
 of the subroutine we  end up with the state  $|A\rangle$ if the
 outcome is 0 and the state $|B\rangle$ if the outcome is 1. The
 parameter $\alpha$ can and should be tuned a little, such that this
 will be an exact and not an approximate statement. Exact, except
 for a possibility of the failure which can be made arbitrary small,
 since it is of the order of $1\over N$. The whole procedure ends with
the test in which cavity, $A$ or $B$, the photon is located, and
this yields the outcome of the computation, 0 or 1.

It seems that it is a counterfactual computation. Indeed, if the
outcome is 1, we know that the photon was not inside the computer
(otherwise it would be absorbed), and, if the outcome is 0, we also
apparently can claim that the photon was not inside the computer,
because our tests of the cavity $C$ at the end of each subroutine
checked every time when the photon could enter the computer, and we
found that it did not.

The core of the controversy is  that it is possible to make CFC when
the process of computation is just the passage of the photon through
the device without any change in the device and without absorption
of the photon; it corresponds to the CFC of the outcome 0 in the
examples above. So, following Hosten et al. \cite{HoQ}, let us
consider a simpler scheme. It is not optimally efficient and it
makes decisive computation only in the case of the outcome 0.

The scheme consists of one Mach-Zehnder interferometer (MZI) nested
inside another, and the computer placed in one arm of the inner
interferometer, see Fig. 3. The inner interferometer is tuned by a
$\pi$ phase shifter in such a way that if the outcome is 0, i.e.,
the computer is transparent, there is a destructive interference
towards the output beam splitter of the large interferometer. If the
outcome is 1, the photon can reach the output beamsplitter of the
external MZI, and it is tuned in such a way that there is a
destructive interference toward detector $D_1$.
\begin{figure}[b]
\includegraphics[width=6.9cm]{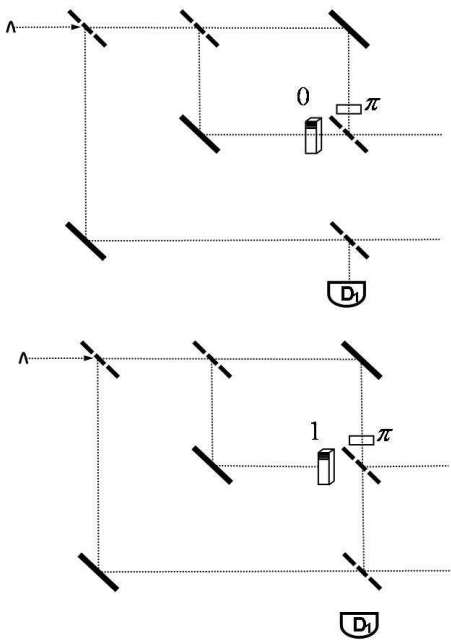}\\
\caption{The Hosten et al.  counterfactual computation method for
outcome 0. The first figure  shows that if the outcome is 0, then
the photon entering the inner interferometer cannot reach detector
$D_1$, and the second figure shows that if the outcome is 1, then no
photon at all  can reach detector $D_1$. } \label{3}
\end{figure}

A possible implementation of this system is that the external MZI
has two identical beamsplitters which transmit two thirds of the
beam and reflect one third. The evolution of the photon passing
through such a beamsplitter is:
 \begin{equation}
\label{U1}\nonumber |H\rangle \rightarrow  \sqrt{2\over 3} |H\rangle
+ \sqrt{1\over 3} |V\rangle ,  ~~
 |V\rangle \rightarrow  -\sqrt{1\over 3}  |H\rangle + \sqrt{2\over 3}
 |V\rangle,
\end{equation}
where $H$ and $V$ signify horizontal and vertical modes. The inner
MZI has two identical half and half beamsplitters. The evolution of
the photon passing through such beamsplitter is:
\begin{equation}
 \label{U2}\nonumber
|H\rangle \rightarrow  \sqrt{1\over 2}  |H\rangle + \sqrt{1\over 2}
|V\rangle,~~
 |V\rangle \rightarrow  -\sqrt{1\over 2}  |H\rangle + \sqrt{1\over 2}
 |V\rangle.
 \end{equation}

A simple calculation shows that, indeed, there is a destructive
interference  toward vertical output mode of the inner
interferometer if the computer is transparent, and there is a
destructive interference toward $D_1$ if the lower arm of the inner
interferometer is blocked by the computer.

Consider now a run of our device in which a single photon enters the
interferometer and it is detected by detector $D_1$. In this case we
get the information that the result of the computation is 0 and it
is apparently a counterfactual computation. It seems that the photon
has not passed through the computer since photons passing through
the inner interferometer, where the computer is located, cannot
reach detector $D_1$.

In spite of the fact that it is a very vivid and persuasive
explanation, I will argue that it is incorrect. The photon detected
at outport $D_1$ is a quantum pre- and post-selected system, and
what is correct for classical systems and quantum pre-selected only
systems might be wrong for a pre- and post-selected system.

Let us consider an example of a pre- and post-selected systems,
usually known as  the ``three-box paradox'' \cite{AV91}. A single
particle is pre-selected in the state
\begin{eqnarray} \left|\psi\right\rangle  & = &
\frac{1}{\sqrt{3}}\left(\left|A\right\rangle +\left|B\right\rangle
+\left|C\right\rangle \right)\,\,\,,\label{eq:3boxpre}\end{eqnarray}
 and post-selected in the state
  \begin{eqnarray}
\left\langle \phi\right| & = & \frac{1}{\sqrt{3}}\left(\left\langle
A\right|+\left\langle B\right|-\left\langle
C\right|\right)\,\,\,,\label{eq:3boxpost}
\end{eqnarray}
where the mutually orthogonal states $\left|A\right\rangle $,
$\left|B\right\rangle $, and $\left|C\right\rangle $ denote the
particle being in box $A$, $B$, and $C$, respectively. (We use
``bra'' and ``ket'' notation to distinguish between standard,
forward evolving quantum state and backward evolving quantum state
from the post-selection measurement.)

The paradox is that at the intermediate time, the particle is to be
found with certainty in box $A$ if searched there and, at the same
time, it is to be found with certainty in box $B$ if it is searched
there instead. Indeed, if a particle is not found, e.g., in box $B$,
then its state collapses to
$\frac{1}{\sqrt{2}}\left(\left|A\right\rangle +\left|C\right\rangle
\right)$, but this is impossible since this state is orthogonal to
the post-selected state (\ref{eq:3boxpost}).

 One thing is obvious in this
example: there are no grounds to say that the particle was not
present at box $B$.  Nevertheless, I can ``show'' that it was not
present in $B$  using similar arguments to those showing that the
photon was not passing through the computer in the Hosten et al.
example.

In fact, the Hosten et al. setup is  an implementation of the
three-box experiment. Consider three boxes on the way of the photon
in the three arms of the nested MZI experiment, Fig. 4. Taking into
account the evolution laws (\ref{U1},\ref{U2}) we see that indeed,
quantum state of the photon in these boxes is described by
(\ref{eq:3boxpre}), and, given that the computer is transparent,
detection by $D_1$ corresponds to post-selection of the state
(\ref{eq:3boxpost}).
\begin{figure}[t]
\includegraphics[width=7.6cm]{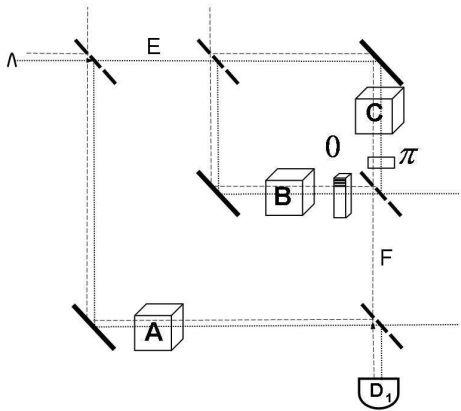}\\
\caption{Both forward and backward evolving quantum states are
present in all three boxes: $A$, $B$, and $C$; they are described by
(\ref{eq:3boxpre}) and (\ref{eq:3boxpost}). However, on the way
toward inner interferometer, backward evolving wave is not present
and on the way out, the forward evolving wave is not present. }
\label{4}
\end{figure}

According to Hosten et al.,  box $B$ is empty.  But this
argumentation is not tenable. Indeed, their argument applies also to
box $C$, so it should be empty too. The pre- and post-selection
states are  symmetric under interchange of boxes $A$ and $B$.
Therefore, box $A$ should be empty as well, but then: Where is the
particle? This shows that the Hosten et al. argument leads to a
contradiction and therefore their conclusion regarding the
counterfactual nature of computation is not warranted.

Let us now ask: What can we learn from  experiments testing location
of the photon in the Hosten et al. setup?  We know that the photon
is to be found with certainty if searched in $B$. One can argue,
however, that a strong nondemolition measurement of the projection
on $B$ changes the physical situation. Then, we can perform a {\it
weak measurement} \cite{AV90} of the projection on $B$ which
requires an ensemble of $N$ pre- and post-selected particles. The
weak measurement, at the limit of large $N$, does not change neither
the forward evolving state (\ref{eq:3boxpre}), nor the backward
evolving state (\ref{eq:3boxpost}).  According to the Hosten et al.
argument, all members of the ensemble are not in $B$, so we should
not see any effect in the measurement at $B$. The experiment,
however, will show a different result: In the CFC experiment the
{\em weak value}, the outcome of the weak measurement of the
projection onto the ``computer'' is 1, while simultaneous weak
measurements of the projections on the paths $E$ and $F$, which lead
to and from the inner interferometer, show 0. This is not
necessarily a gedanken experiment. See experimental results of weak
measurement of the three-box problem, Resch et al. \cite{3box-exp}.
In the framework of these concepts we can state:
 The photon did not enter the interferometer, the photon
never left the interferometer, but it was there! This is a new
paradoxical feature of a pre- and post-selected quantum particle.

Hosten and Kwiat \cite{HK} posed a question about the small, yet
unavoidable disturbance due to weak measurements. To which extend
the weak measurement at $B$ disturbs the destructive interference in
$F$, which is the basis of the Hosten et al. argument? If we perform
a practical weak measurement procedure ala Resch et al., in which
each photon has its own measuring device (its transversal location),
then finding precise weak value requires strength of the interaction
proportional to $1\over \sqrt N$, and thus, a flux through $F$ of a
number of photons. Still, the flux through $F$ is negligible compare
to the total flux, so that the destructive interference in $F$ for
every photon is almost complete.  Moreover, for the conceptual issue
discussed here, we can consider a gedanken experiment in which we
use an external measuring device interacting very weakly with all
the photons and rare quantum event in which all photons are
post-selected in $D_1$. In this experiment a precise weak value of
the projection on $B$ can be obtained with interaction strength
 proportional to $1\over  N$ and thus, a flux through $F$ is much
less than one photon. The operational meaning of this statement is
that strong nondemolition measurements along path $F$ have
negligible probability to find even one photon during the whole
process. It should be mentioned that the disturbing effect of the
strong measurement at $F$ on weak measurement at $B$ is not
negligible at all: Although the measurement at $F$ does not change
the forward evolving quantum state at $B$, it nullifies the backward
evolving state causing the weak value of the projection on $B$ to
vanish.

Note, that Bohmian interpretation of quantum mechanics does not
exhibit such a behavior and, in fact, supports the claim of Hosten
et al. A Bohmian particle has to ``ride'' on a quantum wave, but
there is no quantum wave at path $F$, from the inner interferometer
to $D_1$. Therefore, the (Bohmian) particle detected by $D_1$ did
not pass through the interferometer and the computer.

Discarding the hidden variables approach, we should be able to
answer any question based on the complete description of a quantum
system which consists here of both forward and backward evolving
quantum states. The answer cannot depend on the particular way of
preparing and post-selecting these states, as it happens to be if we
adopt the Hosten et al. approach.

It is a pleasure to thank Zhe-Xuan Gong for helpful discussions.
This work has been supported in part by the European Commission
under the Integrated Project Qubit Applications (QAP) funded by the
IST directorate as Contract Number 015848 and by grant 990/06 of the
Israel Science Foundation.


\end{document}